\documentclass[useAMS,usenatbib]{mn2e}
\usepackage{epsfig}

\def\msun{{\rm {M}}_{\odot}}
\newcommand{\etal}{{et al.}~}
\newcommand{\eg}{{e.g.~}}
\newcommand{\ie}{{i.e.~}}
\newcommand{\apj}{{ApJ}}

\def \ltsima{$\; \buildrel < \over \sim \;$}
\def \simlt{\lower.5ex\hbox{\ltsima}}            % < over ~
\def \gtsima{$\; \buildrel > \over \sim \;$}
\def \gtsima{\mbox{$\; \buildrel > \over \sim \;$}}
\def \simgt{\lower.5ex\hbox{\gtsima}}            % > over ~
\title[clustering and ages]{The age dependence of galaxy clustering}
\author[Reed \etal] {Darren S. Reed,$^{1,2}$\thanks{Email:
reed@lanl.gov}
Fabio Governato,$^{3,4}$
Thomas Quinn,$^3$
\newauthor
Joachim Stadel,$^5$
and George Lake$^5$\\
$^1$Institute for Computational Cosmology, Dept. of Physics, University of
Durham, South Road, Durham DH1 3LE, UK\\
$^2$Theoretical Astrophysics Group, Los Alamos National Laboratory, 
PO Box 1663, MS 627, Los Alamos, NM, 87545 USA\\
$^3$Astronomy Department, Box 351580, University of Washington, Seattle,
WA 98195 USA\\
$^4$INAF, Osservatorio Astronomico di Brera, via Brera 28, I-20131 Milano,
Italy\\
$^5$Institute for Theoretical Physics, University of Zurich,
Winterthurerstrasse 190, 8057, Switzerland}
\pagerange{\pageref{firstpage}--\pageref{lastpage}}
\pubyear{2004}

\begin{document}

\maketitle

\label{firstpage}

\begin{abstract}

We construct mock galaxy catalogues to analyse clustering properties
of a $\Lambda$ cold dark matter ($\Lambda$CDM) universe within a
cosmological dark matter simulation of sufficient resolution to
resolve structure down to the scale of dwarfs.  We show that there
is a strong age-clustering correlation for objects likely to host luminous
galaxies, which includes the satellite halo (subhalo) population.
Older mock galaxies are significantly more clustered in our catalog,
which consists of satellite haloes as well as the central peaks of
discrete haloes, selected solely by peak circular velocity.  
This age dependence is caused mainly by the 
age-clustering relation for discrete haloes, recently found by Gao \etal, 
acting mostly on field members, combined with the tendency for older 
mock galaxies to lie within groups and clusters, where galaxy 
clustering is enhanced.  
Our results 
suggest that the clustering age dependence is manifested in real
galaxies.  At small scales (less than $\sim$5 $h^{-1}$Mpc), the very
simple assumption that galaxy colour depends solely on halo age is
inconsistent with the strength of the observed clustering colour
trends, where red galaxies become increasingly more clustered than
blue galaxies toward smaller scales, suggesting that luminosity
weighted galaxy ages do not closely trace the assembly epoch of their
dark matter hosts.  The age dependence is present but is much weaker for
satellite haloes lying within groups and clusters than for the global
population.

\end{abstract}

\begin{keywords} galaxies: haloes -- galaxies: formation -- methods:
N-body simulations -- cosmology: theory -- cosmology:dark matter
\end{keywords}

\section{introduction}

A critical test of the $\Lambda$CDM model is whether it
accurately predicts the clustering properties of galaxies formed
within dark matter haloes and ``subhaloes'' that are satellite clumps
withing larger host haloes.  Subhaloes serve as hosts for visible
galaxies within clusters, groups, or larger galaxies, and so provide a
natural basis for constructing simulated mock catalogs, whose
clustering properties can then be compared with observed galaxies.

Clustering of haloes depends on halo age, a phenomenon recently
measured in $\Lambda$CDM simulations by Gao, Springel, \& White
(2005), who found that older haloes are more strongly clustered than
younger haloes.  The likely explanation is that haloes of a given mass
generally form earlier within denser regions.  Thus, older haloes tend
to populate denser regions, which naturally leads to stronger
clustering with halo age.  
Sheth \& Tormen (2004) measured such a
trend between mean halo formation epoch and local over-density in
simulation data, which was recently confirmed in larger simulations by
Harker \etal (2005).
The precise physical origin, however, of the age dependence is a
subject of recent debate (see discussion by \eg Zentner 2006).
Wang, Mo \& Jing (2007), use numerical simulations to suggest 
that accretion onto low mass haloes in high density regions
is inhibited by competition
with massive neighbours via tidal interactions and local dynamical heating,
creating a correlation with halo age and environment.
Sandvik \etal (2006) suggest that the formation history and the 
current epoch environment of low mass haloes may be affected by 
their presence in massive pancakes and filaments at high redshift. 
The clustering age dependence, which has been
confirmed by a number of authors (Harker \etal 2005; 
Wechsler \etal 2006; Gao \& White 2006; Jing, Suto \& Mo 2006; Wang, Mo \& 
Jing 2007), is strong
for haloes of masses of 10$^{11-12}$ $h^{-1} \msun$ (Gao \etal 2005,
Wechsler \etal 2006), which are likely to host galaxies, 
and decreases with halo
mass, becoming insignificant for haloes more massive than
$\sim$10$^{13}$ $h^{-1} \msun$ (Gao \etal 2005). There
is evidence that the age-clustering dependence may
reverse
sign for larger haloes (Wetzel \etal 2006; Jing, Suto \& Mo 2006;
Gao \& White 2006; Zentner 2006).

Previous studies have focussed on the age dependence of clustering of
discrete virialized haloes, and did not consider directly the
contribution of satellite populations to the age dependence of
clustering.  Because a large fraction of galaxies belongs to groups
and clusters, the clustering of the general galaxy population could
have a strong dependence on subhalo ages.  
Galaxies of a given circular velocity will have formed
earlier if they lie in groups or clusters today.  Thus, we can
expect that the contribution of group and cluster members will
increase the tendency for older objects to be more strongly
clustered. Also, recent studies relating
subhalo numbers and distribution to age and to host halo properties
hint that subhalo clustering could depend on subhalo age.  Recent
simulations have shown that older haloes tend to host fewer subhaloes
(\eg Gao \etal 2004; Zentner \etal 2005; Taylor \& Babul 2005; Zhu
\etal 2006).  Additionally, the clustering strength of virialized
haloes is correlated with the numbers of their satellite haloes
(Wechsler \etal 2006).  Furthermore, subhaloes tend to lie nearer
their host centres if they were either formed earlier (Willman \etal
2004) or were accreted earlier (Gao \etal 2004; Taylor \& Babul 2005;
see however Moore, Diemand, \& Stadel 2004).

In order to understand more fully the age dependence of subhalo
clustering and its potential effects on observable galaxies, we
analyze the relation between age and clustering within halo
catalogs that include both the satellite haloes that populate group
and cluster haloes as well as the discrete virialized haloes likely to
host only a single galaxy.  We construct a simple mock galaxy catalog
wherein haloes are selected by peak circular velocity to roughly match
the galaxy luminosities and abundances in large surveys.  Our
catalog is selected from a high resolution dark matter simulation that
resolves structures within a cosmological volume down to the scale of
dwarf galaxies.  We stress that we are not attempting to create a
realistic catalog of ``simulated galaxies'', but rather that we are
merely using observationally relevant circular velocities as a
convenient means of assessing the potential dependence of clustering
on age of haloes$+$subhaloes over a range that has the potential to
host galaxies in the $\Lambda$CDM model.  In \S~2, we describe the
simulations and the construction of the halo$+$subhalo catalog.  In
\S~3, we detail the age dependence of clustering in our mock catalog,
the implications of which we discuss in \S~4.

\section{numerical techniques}
\subsection{the simulations}

We use the parallel k-D (balanced binary) Tree (Bentley 1975) gravity
solver {\small PKDGRAV} 
(Stadel 2001; Wadsley, Stadel \& Quinn 2004) to model a
50 $h^{-1}$Mpc cube, consisting of 432$^{3}$ dark matter particles of
equal mass (the CUBEHI run of Reed \etal (2003; 2005ab).  By modelling
a relatively small cosmological volume, we are able to probe down to
the small masses needed to resolve satellites within groups.  The
particle mass is 1.3$\times 10^{8} h^{-1} \msun$.  A starting redshift
of 69 and a force softening of 5 $h^{-1}$kpc (comoving) are used.  This
run adopts a $\Lambda$CDM cosmology with $\Omega_m=$ 0.3 and
$\Lambda=$ 0.7, and the initial density power spectrum is normalised
to $\sigma_{\rm 8}=$ 1.0, consistent with WMAP (\eg Bennett \etal
2003; Spergel et al. 2003). We use a Hubble constant of $h=0.7$, in
units of 100 km s$^{-1}$ Mpc$^{-1}$, and assume no tilt (i.e. a
primordial spectral index of 1). To set the initial conditions, we use
the Bardeen \etal (1986) transfer function with $\Gamma=\Omega_{\rm
m}\times h$.

\subsection{mock catalog construction} 

Mock galaxies are chosen from a catalog of haloes selected by circular
velocity using the Spline Kernel Interpolative {\small DENMAX} ({\small SKID}) halo
finder (Stadel 2001;
http://www-hpcc.astro.washington.edu/tools/skid.html).  {\small SKID} haloes
are identified using local density maxima to identify bound mass
concentrations independently of environment.  Note that {\small SKID}
identifies discrete virialized haloes as well as subhaloes (satellite
haloes).  The radial extent of each {\small SKID} halo is determined by the
distribution of bound particles, and no predetermined subhalo shape is
imposed.  The peak circular velocity of each subhalo, $v_{c,peak}$, is
computed from the peak of the rotation curve ${\rm v_c(r) =
(GM(<r)/r)^{0.5}}$.  The formation epoch is defined as the time at
which $v_{c,peak}$ of the largest progenitor (amongst all
branches of the merger tree at a given epoch) reaches 75$\%$ of its
maximum value.  A progenitor is defined as a halo with at least
30$\%$ of its particles incorporated into its descendent.  Further
detail on formation and accretion times of subhaloes can be found in a
number of prior studies (\eg De Lucia \etal 2004; Gao \etal 2004).

The mock galaxies are selected to have a magnitude range similar to
that of the Sloan Digital Sky Survey (SDSS) sample analyzed by Zehavi
\etal (2002), $-22 > M_r > -19$, though our results are not sensitive
to the precise range.  The absolute r-band magnitude $M_r$ of each
{\small SKID} halo is estimated by applying $v_{c,peak}$ to the Tully \& Pierce
(2000) variant of the Tully-Fisher (Tully \& Fisher 1977) relation:
\begin{equation}
M_R = -21.12 - 7.65 (log W_R - 2.5),
\end{equation}
where the linewidth $W_r$ is approximately twice $v_{c,peak}$ 
(Tully \& Fouque 1985).  We select approximately 6,000 mock
``galaxies'' with $84~km~s^{-1} < v_{c,peak} < 206~km~s^{-1}$.  In
practice, our faintest simulated galaxies are {\small SKID} haloes of several
hundred particles.  While the Tully-Fisher magnitude assignment is
subject to a number of uncertainties, including the fact that we apply
this to ellipticals as well as spirals (see \eg Desai \etal 2004), it
provides a convenient method for building a catalog of mock galaxies
with magnitudes comparable to those in galaxy surveys.  The spatial
abundance of the mock catalog is 4.8 $\times$ 10$^{-2}$
$h^{3}$Mpc$^{-3}$, which is 2.6 times that of the SDSS sample selected
from the same magnitude range.  However, we stress that our results
are not sensitive to the abundance or to the limits used for inclusion
into the mock catalog; \ie, we are able to detect a
clustering-age dependence for a range of $v_{c,peak}$-selected
catalogs in addition to the one presented here, as we show later.
Thus, even though the
objects in our catalog are not expected to describe precisely the
galaxy population, we can still capture many of the important
clustering properties of the dark hosts of galaxies. 

\subsection{correlation functions}
 
The spatial pairwise correlation function of galaxies is an important
cosmological test, as it quantitatively measures basic clustering
properties (see \eg Peebles 1980).  The spatial correlation function
is calculated using the direct estimator (as in \eg Governato \etal
1999):
\begin{equation}
\xi(r)={2N_p(r)\over n_{c}^2 V (\delta V)} -1,
\end{equation}
where $N_p(r)$ is the number of pairs in radial bins of volume $\delta
V$, centred at $r$; $n_c$ is the mean space density of the catalog;
and $V$ is the volume of the simulation.  We take into account our
periodic boundary conditions when finding pairs.  The correlation
function is often approximated by a simple power law:
\begin{equation}
\xi(r)=\left({r\over r_0}\right)^{-\gamma},
\end{equation}
with $\xi(r_0) = 1$, where $r_0$ is the correlation length.  The
relative clustering amplitude between haloes and the mass distribution
is referred to as bias:
\begin{equation}
b^2 = {\xi_{halo-halo}(M, r, z) \over \xi_{dm}(r ,z))}.
\end{equation}
For all error estimates, we use 1$\sigma$ poisson errors 
(equal to the square root of the number of pairs in each bin), which
are likely to underestimate the true errors because they do not take
into account clustering and sample variance (\eg Croft \&
Efstathiou 1994).  However, because we are interested
primarily in the relative clustering between age-selected
objects, and not the true clustering strength, poisson errors are
adequate for this study.

\section{results}
\subsection{correlation functions of mass and haloes}
 
\begin{figure}
  \begin{center}
  \epsfig{file=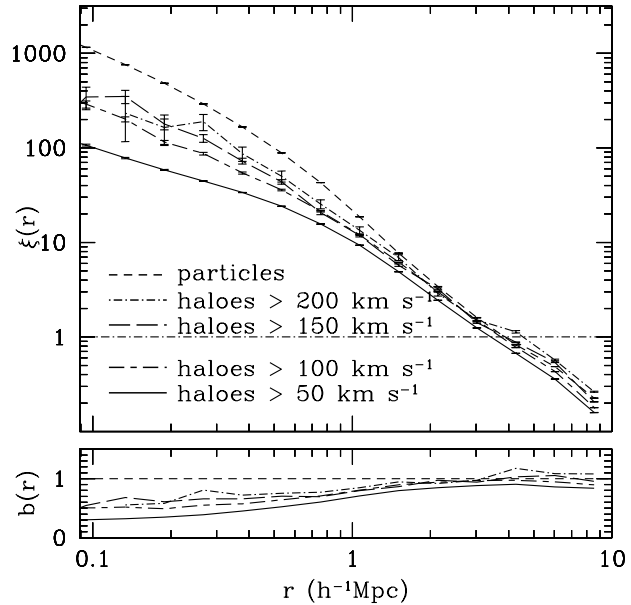, width=\hsize}
  \caption{ {\it Top panel:} The real space two-point correlation
function for our {\small SKID} haloes with ${v_{c,peak} > 50, 100, 150}$, and
$200~km~s^{-1}$, plotted along with $\xi{(r)}$ for particles (mass).  The
horizontal line is $\xi{(r)}=1$.  {\it Bottom panel:} The halo bias
$b(r) \equiv \sqrt{\xi_{haloes}(r)/\xi_{mass}(r)}$.  {\small SKID} haloes are
selected by local density maxima, independent of environment, and
include the centres of discrete virialized haloes in addition to
self-bound satellite haloes (subhaloes).}
\label{cfhostvc}
\end{center}
\end{figure}

In Fig. \ref{cfhostvc}, we plot $\xi(r)$ and the bias factor
($b(r) \equiv \sqrt{\xi_{haloes}(r)/\xi_{mass}(r)}$) for {\small SKID} haloes
with $v_{c,peak} >$ 50, 100, 150, and 200 ${\rm km~s^{-1}}$.  Larger
{\small SKID} haloes have steeper correlation functions and larger correlation
lengths.  The largest {\small SKID} haloes are ``antibiased'' ($b<1$)
with respect to the mass on small scales, but on large scales they are
slightly more clustered than the mass.  In general, haloes are
``antibiased'' with respect to the mass, particularly on small scales
for small haloes.  This is consistent with small scale ``antibias''
found in previous simulations (\eg Jenkins \etal 1998; Colin \etal
1999; Kravtsov \& Klypin 1999; Yoshikawa \etal 2001; Diemand \etal
2004; Reed \etal 2005b) and may be caused by merging or destruction of
subhaloes in high density regions (\eg Jenkins \etal 1998; Klypin \&
Kravtsov 1999).  However, previous similar studies have found
that the correlation function slope does not become shallower at small
scales (Colin \etal 1999; Kravtsov \etal 2004; Neyrinck, Hamilton \&
Gnedin 2004; Conroy, Wechsler \& Kravtsov 2006).  The reasons for the
difference are not clear, but we we note that the correlation function
is a combination of a number of non-powerlaw components
(central-satellite, central-central, and satellite-satellite), so
there is no {\it a priori} expectation that $\xi(r)$ should follow a
power-law (\eg Benson \etal 2000; Berlind \& Weinberg 2002; Kravtsov
\etal 2004).  In fact, the small scale departure from a power law that
we see begins approximately where the satellite-satellite term begins
to dominate $\xi(r)$.  At these scales, $\xi(r)$ could be sensitive to
a number of issues that affect the relative contributions of these
components.  For example, the number of massive clusters, which can
dominate the satellite-satellite term, can be affected by run to run
``sample variance'' or box size (see \eg Reed \etal 2007 and
references therein).  Differences in halo finder behavior at small
scales could also be important.  Further study is warranted, though
our conclusions are not dependent on the smallest scales.  
It is difficult to quantify the precise scale below which the
correlation function will no longer be robust.  However, Reed \etal
(2005b) indicate that the subhalo distribution is robust down to 100
$h^{-1}$kpc for simulations of similar resolution.  For this reason,
we have plotted all correlation functions only down to 100 $h^{-1}$kpc.

\subsection{the age dependence of the mock galaxy catalog correlation function}

\begin{figure}
  \begin{center}
  \epsfig{file=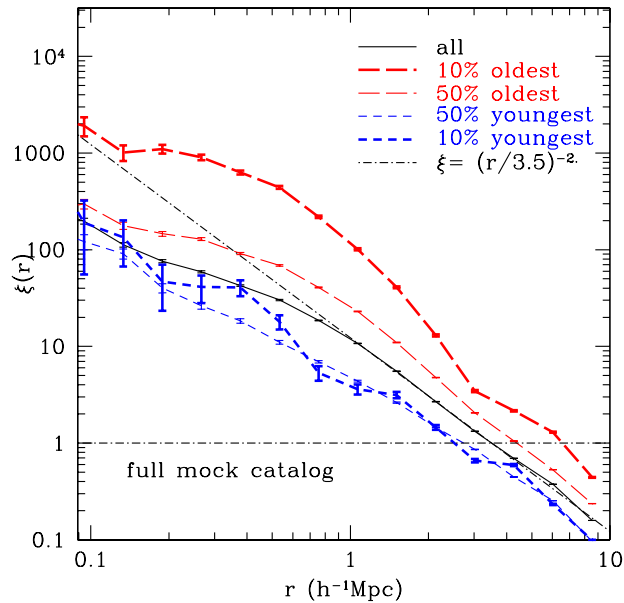, width=\hsize}
  \caption{ The real space two-point correlation function $\xi(r)$ for
our mock catalog, which consists of $v_{c,peak}$-selected {\small SKID} haloes
sorted by age, determined by the time when the circular velocity,
v$_{c,peak}$, of the largest progenitor reaches 75$\%$ of its maximum
value.  The dot-dashed lines corresponds to $\xi(r)=1$, and
$\xi(r)=\left({r\over 3.5}\right)^{-2}$.}
\label{cfvcages}
\end{center}
\end{figure}

\begin{figure*}
  \begin{center}

    \epsfig{file=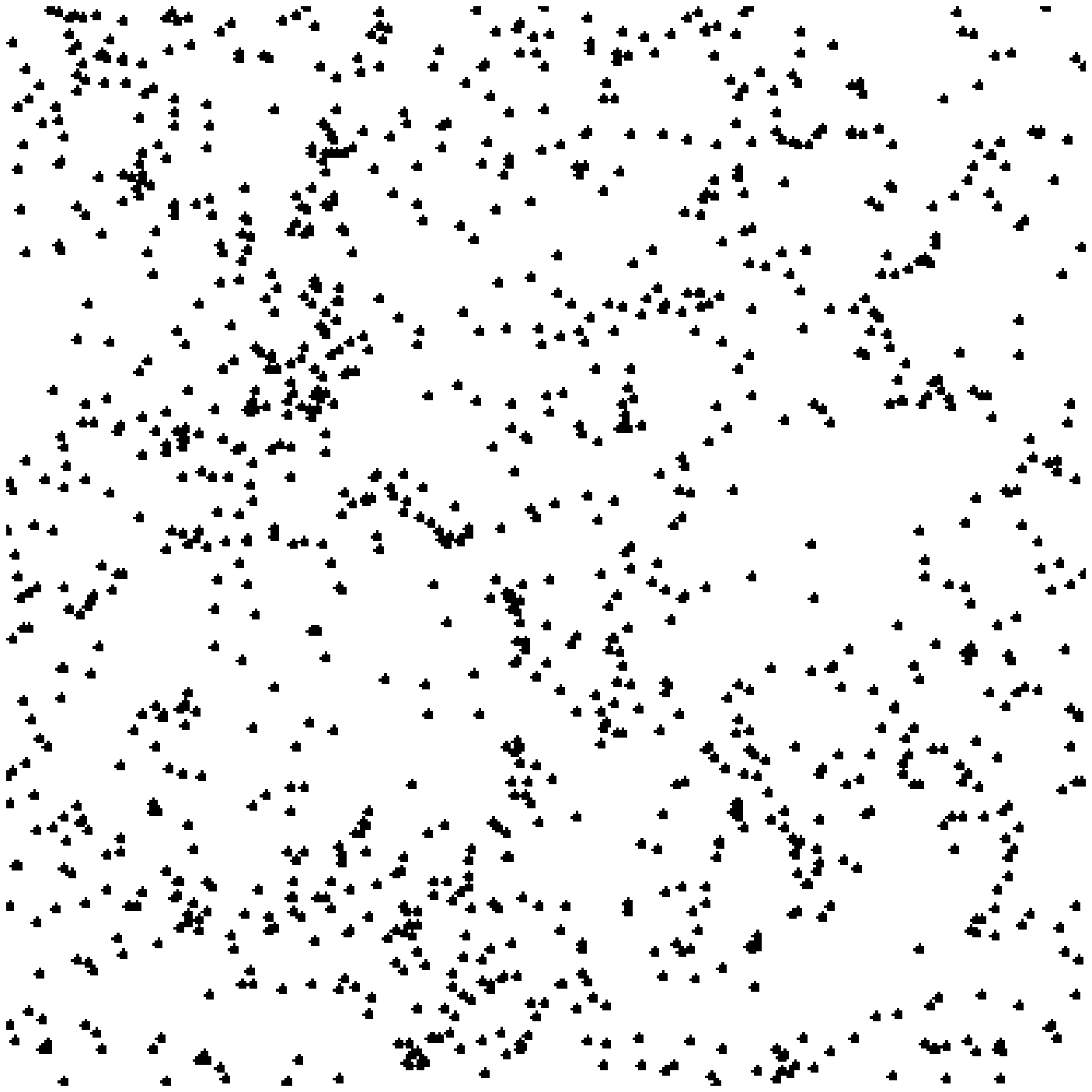, width=0.33\textwidth}
    \epsfig{file=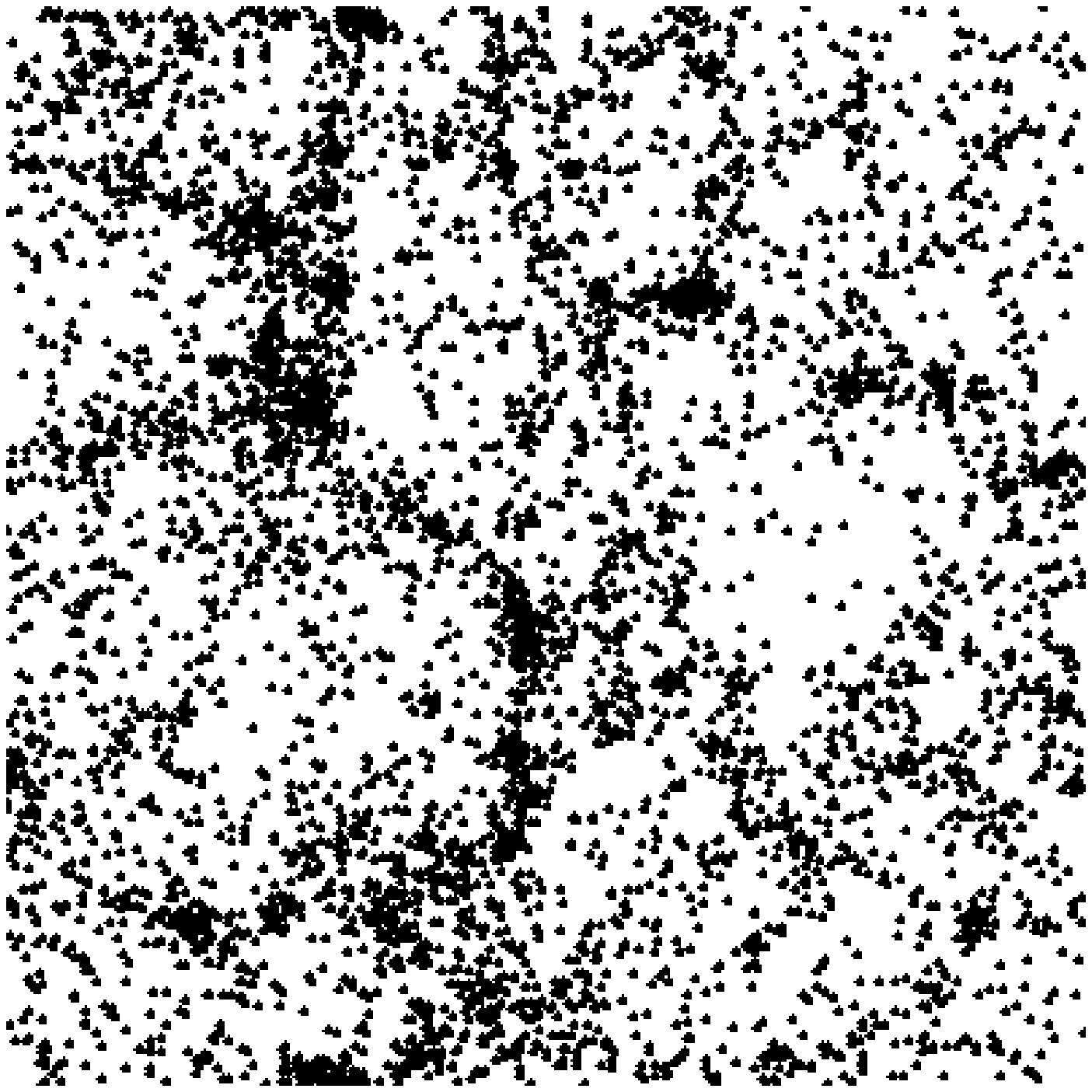, width=0.33\textwidth}
    \epsfig{file=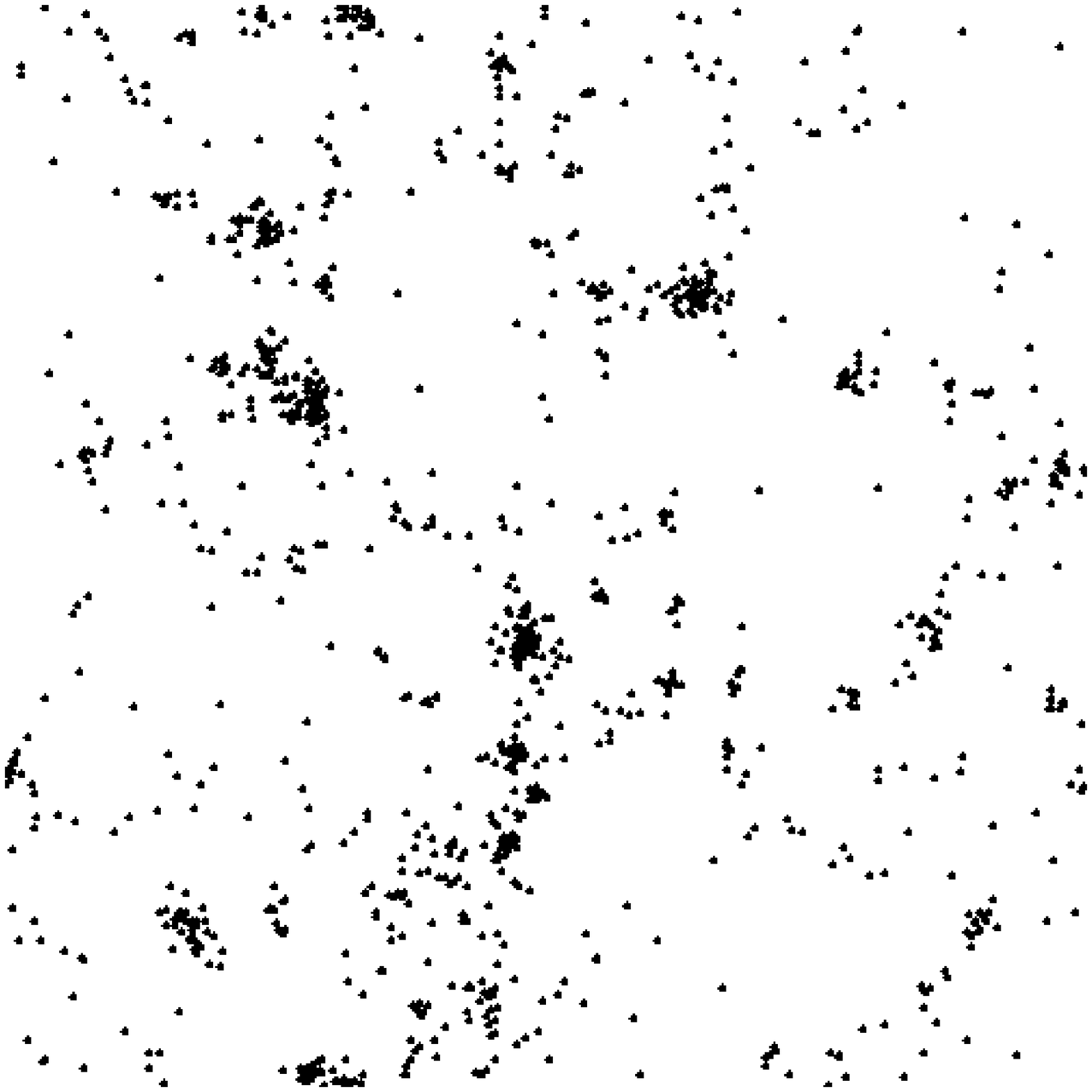, width=0.33\textwidth}

$\begin{array}{c@{\hspace{0.15\textwidth}}c @{\hspace{0.2\textwidth}}c}
\mbox{\bf (a) youngest 20$\%$} & \mbox{\bf (b) full sample} & \mbox{\bf (c) oldest 20$\%$}
\end{array}$

  \caption{ Snapshot shows the mock catalog, composed of
$v_{c,peak}$-selected {\small SKID} haloes, divided into subsets by age,
youngest 20$\%$ (left), full sample (centre), oldest 20$\%$ (right).
Each snapshot shows the entire 50 $h^{-1}$Mpc volume.}
\label{agepics}
\end{center}
\end{figure*}

We plot $\xi{(r)}$ for our mock galaxy sample in Fig. \ref{cfvcages},
binned according to formation times.  Older catalog members are
significantly more clustered for all pair separations.  The
oldest 10$\%$ is most preferentially clustered at small scales, with a
clustering amplitude of $\sim$10$\times$ that of the full mock catalog
for separations less than $\sim$1 $h^{-1}$Mpc.  The differences
between the clustering of the young samples and the full catalog are
smaller, but are significant.
There is little difference in the spatial
correlations of the youngest 10$\%$ and the youngest 50$\%$.  
The striking visual
appearance of the age-clustering dependence is seen in Fig.
\ref{agepics}, which shows the redshift zero simulation snapshot
divided into the youngest and oldest 20$\%$ subsets of the mock
catalog.

\begin{figure}
  \begin{center}
  \epsfig{file=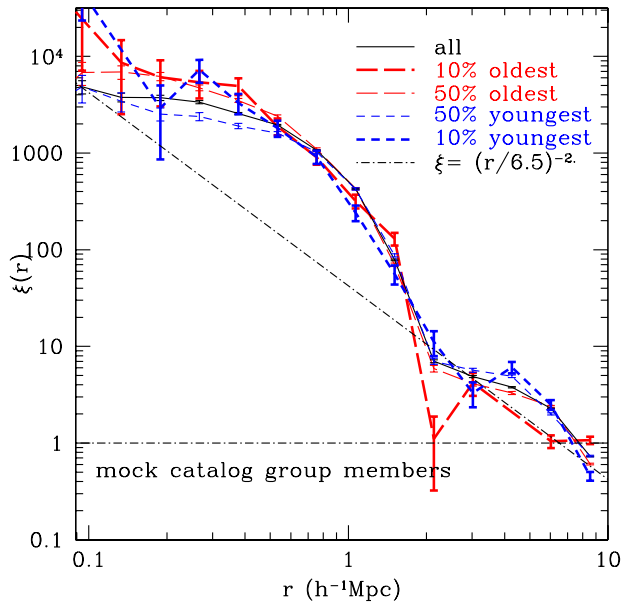, width=\hsize}
  \caption{ Same as Fig. \ref{cfvcages}, but for group members within
the mock catalog, where a group is taken to be an friends-of-friends
identified\ group of at least $3.2 \times 10^{13} \msun h^{-1}$.  
Age percentiles are based on the group members only as opposed to
from the full catalog.
}
\label{cfvcgroupsages}
\end{center}
\end{figure}

\begin{figure}
  \begin{center}
  \epsfig{file=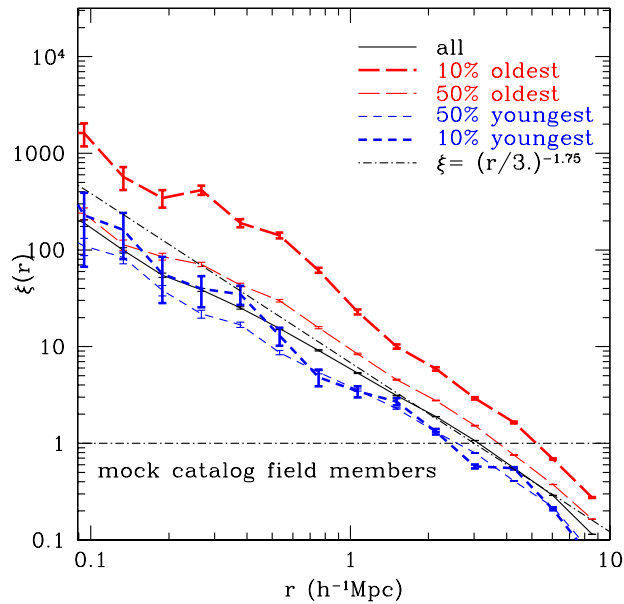, width=\hsize}
  \caption{ Same as Fig. \ref{cfvcages}-\ref{cfvcgroupsages}, but for
field (non-group) members within the mock catalog.  }
\label{cfvcfieldages}
\end{center}
\end{figure}

Fig. \ref{cfvcgroupsages} shows the correlation function for members
of groups or clusters larger than $3.2 \times 10^{13} \msun h^{-1}$.
There is some age dependence of the clustering of group members,  
but it is limited mainly to small
pair separations, and is
significantly weaker than that found for the full mock catalog. 
Clustering of field objects, shown in Fig. \ref{cfvcfieldages}, 
has a strong age dependence, though not as strong found in the complete sample.
To determine group membership, groups are identified using {\it
friends-of-friends} (FOF; Press \& Davis 1982; Davis \etal 1985),
wherein the FOF haloes consist of particles of separated by less than
0.2 times the mean inter-particle separation.  The group extent is
subsequently computed assuming a virial overdensity of approximately
100 times the critical density (Eke, Cole, \& Frenk 1996), and mock
catalog members whose centre of mass lies within this region are
assigned membership to that group.  
The overall correlation amplitude is a 
significantly higher for group members than
for field members, 
an unsurprising result given that group members are
selected deliberately from within regions of high density, and belong 
to massive haloes, which are strongly clustered due to the well-known
mass-clustering relation.

The age-clustering relation in our mock catalog is likely due
to a combination of causes.  For the field sample, the obvious
mechanism is the Gao \etal age-clustering correlation for discrete
virialized haloes.  
For the full field plus group and cluster catalog
(Fig. \ref{cfvcages}), the age dependence is stronger than that of the
field sample alone (Fig. \ref{cfvcfieldages}) because group and
cluster members, which are found in 
highly clustered environments, tend to be old.
Even though group and cluster members comprise only $\sim 10\%$ of the
full sample, their contribution to the clustering age-dependence is
significant due to the strong age correlation with environment.
For example, 80$\%$ of our group and cluster members are older than
the median mock galaxy age; and group and cluster members are 10 times
more likely to belong to the 10$\%$ oldest subset than to the 10$\%$
youngest subset.

On small scales (less than $\sim 1 h^{-1}$Mpc), dynamical interactions 
become important for group and cluster members.  
Upon accretion onto a group or cluster halo, the subhalo will spiral 
in via dynamical friction, undergoing tidal stripping in the process.
This leads to a subhalo distribution where centrally located
group and cluster members were accreted earlier and are older.  This
is the likely cause of the small scale 
age-dependence within the group and cluster subsample.
Finally, for mock galaxies belonging to groups or cluster of 
similar mass, the age dependence of their host group-group clustering
may produce some effect on the
correlation function, but it should be mild because our group and
cluster hosts are larger than the
$\sim10^{13} \msun h^{-1}$ mass threshold
above which the age-clustering relation of discrete haloes becomes
weak (Gao \etal 2005).

\subsection{can the age-clustering relation cause the observed colour-clustering dependence?}
\begin{figure}
\begin{center}
\epsfig{file=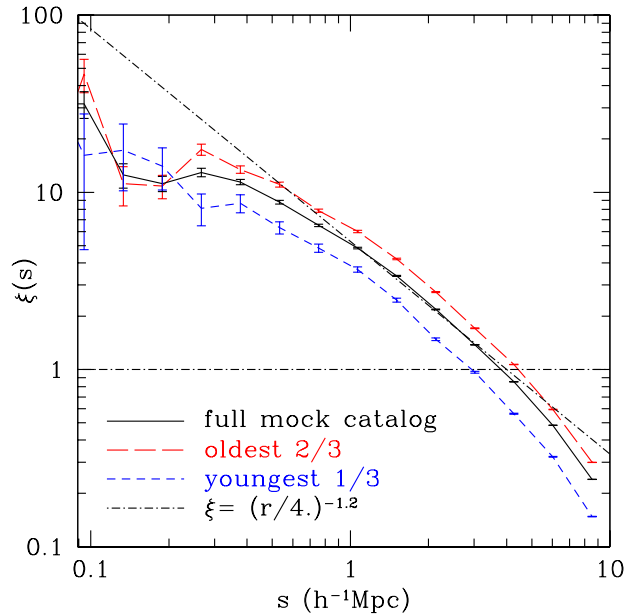, width=\hsize}
\caption{The redshift space two-point correlation function $\xi(r)$
for our mock galaxy catalog of $v_{c,peak}$-selected {\small SKID} haloes
sorted into {\it old} and {\it young} subsamples based on the time at
which the circular velocity of the largest progenitor reaches 75$\%$
of its maximum.}
\label{cfredages}
\end{center}
\end{figure}

In this section, we perform a simple test to determine whether the
magnitude of the age-clustering dependence seen in simulations could
be sufficient to account for the observed colour-clustering dependence
under the simple assumption that halo age is a proxy for galaxy
colour.  Here, we have split the catalog into an ``old'' and a
``young'' subsample with a 2:1 ratio of old to young haloes, which
matches the ratio of red to blue galaxies of the Zehavi \etal (2002)
SDSS sample.  In Fig., \ref{cfredages}, we plot the redshift space
two-point correlation function for these catalog subsamples.  The
magnitude of the age dependence in the mock catalog sample is
comparable to the clustering colour-dependence in SDSS for pair
separations larger than $\sim$5 $h^{-1}$Mpc; see Fig. 11 of Zehavi
\etal (2002) for the SDSS comparison.  However, the age-effect is much
weaker relative to the observed colour trends for smaller separations,
and is insignificant (within the uncertainties) for pairs separated by
200 $h^{-1}$kpc or less.  The observed galaxy clustering
colour-dependence, however, extends down to $\sim$ 100 $h^{-1}$kpc
(\eg Zehavi \etal 2002; 2005, Madgwick \etal 2003; Li \etal 2006).
The redshift space correlation function suggests that the age
dependence has the potential to account for the observed
colour-clustering trends only at large galaxy pair separations.  This
provides an independent argument that galaxy luminosity-weighted ages,
indicated by colour, are different from the ages of the host dark
matter subhaloes in which they lie.

\begin{figure}
\begin{center}
\epsfig{file=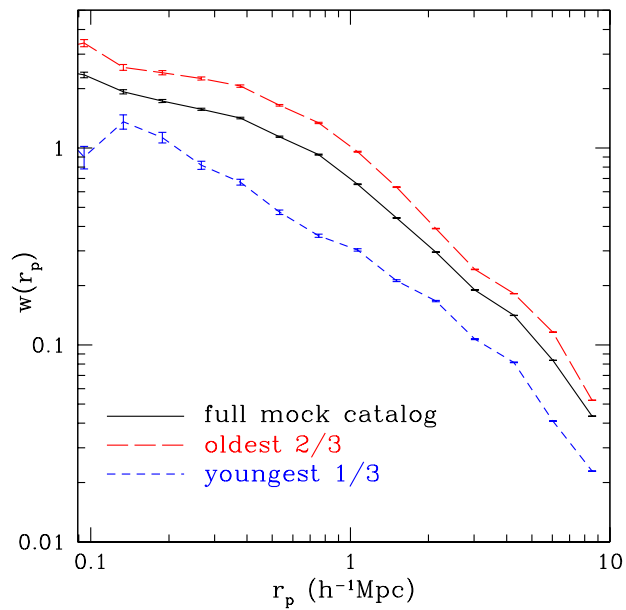, width=\hsize}
\caption{Same as Fig. \ref{cfredages} but showing the
projected two-point correlation function ${\rm w_p(r_p)}$.  }
\label{cfprojages}
\end{center}
\end{figure}

Qualitative differences between age selected simulated haloes
and colour-selected observed haloes are also apparent in the
projected-space two-point correlation function, 
\begin{equation}
w(r_p)={2N_p(r)\over \mu^2 A(\delta A)} -1,
\end{equation}
where ${\rm w(r_p)}$ is the pair excess over random with projected
separation ${\rm r_p}$ binned with area $\delta A$, and $\mu$ is the
mean projected density of haloes over the projected simulation area
$A$ (see \eg Peebles 1980).  In Fig. \ref{cfprojages}, the differences
between the overall slope and small scale amplitude of ${\rm
w_p(r_p)}$ are relatively small between the young and old simulation
samples.  However, observed red galaxies have a much steeper and
larger amplitude ${\rm w_p(r_p)}$ than blue galaxies (see \eg Zehavi
\etal 2002, Fig. 13; Zehavi \etal 2005, Fig. 13) for a wide 
range of luminosity selected samples (\eg Li \etal 2006).

We note that our overall correlation amplitude is significantly
smaller than that of the SDSS sample.  This is due, at least in part,
to our finite box size, which means that large scale density
fluctuations are not fully and accurately represented 
(\eg Bagla \& Ray 2005; Sirko 2005;
Power \& Knebe 2006; Reed \etal 2007), 
and should not be interpreted as an indication
of a conflict with observations.  A further contribution to our lower
clustering amplitude may be the higher spatial abundance of our mock
catalog, which implies that we are selecting smaller, and hence less
strongly clustered, objects than in the SDSS sample.

\begin{figure}
\begin{center}
\epsfig{file=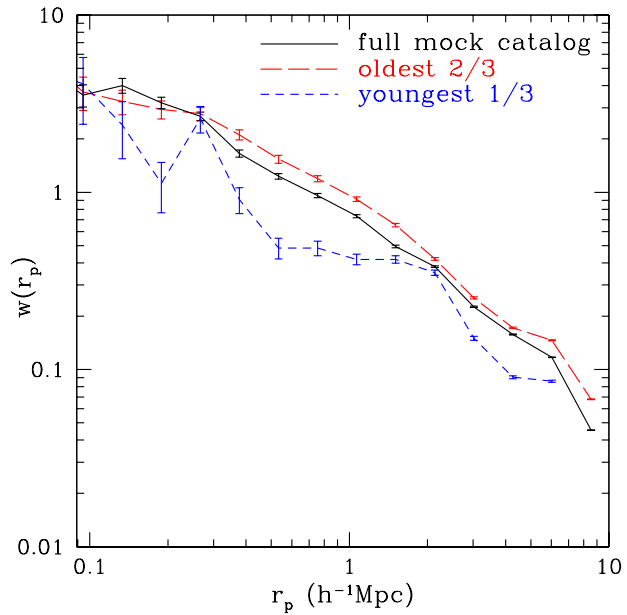, width=\hsize}
\caption{Same as Fig. \ref{cfredages}, but showing 
${\rm w_p(r_p)}$
for a catalog selected by $v_{c,peak} > 150 {\rm km s}^{-1}$.  }
\label{cf150projages}
\end{center}
\end{figure}

A caveat here is that the mock catalog is inherently different
from the SDSS sample.  Because the correlation function of the full mock
catalog is not a power law, as is generally observed in real galaxies, one
should question whether the relative correlations of age-selected
samples will display the same properties as real galaxies.  
A truly realistic simulated galaxy sample would of course require
modelling correctly 
all baryon physics, including star formation and feedback at
small scales, a task that is not feasible at this time.  
However, it is prudent to consider how
the details of our mock catalog construction could affect the 
measured age-clustering signal.
To enable better statistics, our 
mock catalog was selected to contain smaller galaxies than the SDSS sample. 
It is clear from
Fig. \ref{cfhostvc} that a sample of mock galaxies with higher $v_{c,peak}$ 
results in a correlation function that is closer
to a power law, better matching observations.  
We thus make a test to show whether the
age-dependence of the correlation function is sensitive to
$v_{c,peak}$ (or abundance).  In Fig \ref{cf150projages}, we show that
the clustering dependence on age has similar qualitative
behavior, although the age dependence is somewhat weaker, 
for a sample of larger mock galaxies selected purely by 
$v_{c,peak} > ~150 ~km ~s^{-1}$.
This indicates that the scale dependence of age-clustering 
relation in our simulation
is relatively insensitive to our specific choice of $v_{c,peak}$ range.

\subsection{age dependence of the pairwise velocity dispersion}
 
\begin{figure}
  \epsfig{file=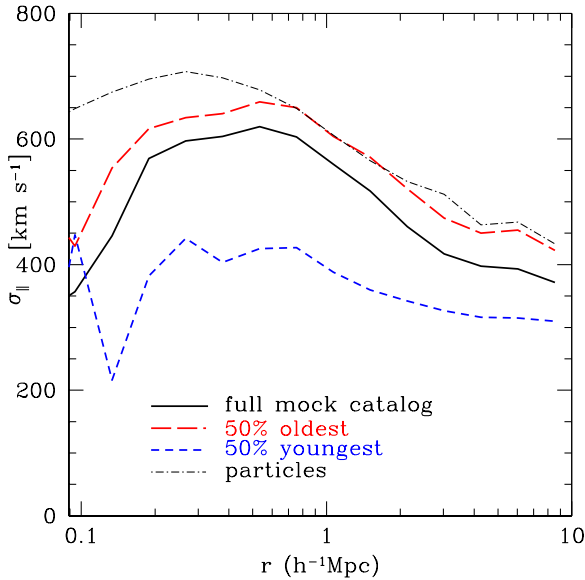, width=0.45\textwidth}
  \caption{The 1-D pairwise velocity dispersion along the line of
separation, $\sigma_{\parallel}(r)$, is plotted for our
$v_{c,peak}$-selected mock catalog, divided into a young and an old
subset.}
\label{cvages}
\end{figure}

Galaxy peculiar velocities, as measured by the pairwise velocity
dispersion, also provide a valuable probe of galaxy clustering, as
well as providing an important component for dynamical probes of the
dark energy equation of state (\eg Governato \etal 1997; Baryshev,
Chernin \& Teerikorpi 2001).  The 1-D pairwise velocity dispersion,
$\sigma_{\parallel}$(r), is the velocity dispersion for particle pairs
in the direction parallel to the line of separation.  In
Fig. \ref{cvages}a, we plot $\sigma_{\parallel}(r)$ for our mock
catalog and for a random subsample of particles.  Old mock galaxies
have substantially higher pairwise velocities than young mock
galaxies, as expected given their higher degree of spatial clustering.
The old galaxy $\sigma_{\parallel}$(r) is $\sim$ 50 $km~s^{-1}$
``hotter'' than the combined sample, and the young sample is
``cooler'' by up to 200 $km~s^{-1}$.  The lower pairwise velocities
for the mock catalog with respect to the dark matter particles is
consistent with its spatial ``antibias'', shown in
Fig. \ref{cfhostvc}.

\subsection{age dependence in a mass-selected sample}

\begin{figure}
  \begin{center}
  \epsfig{file=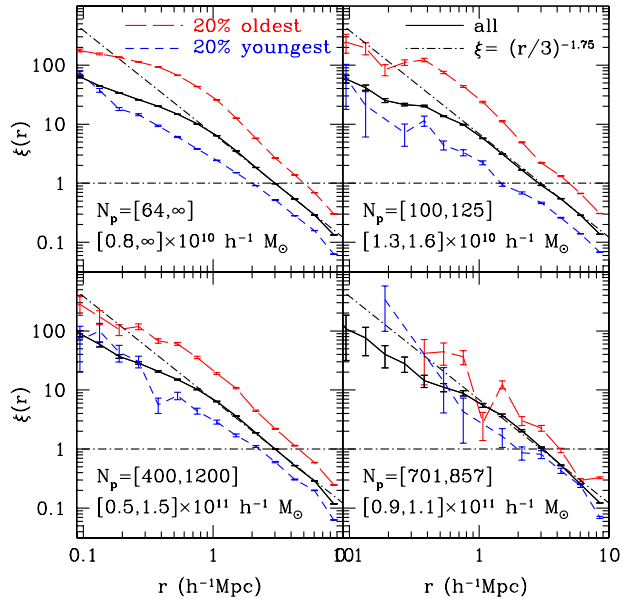, width=\hsize}
  \caption{The two-point correlation function $\xi(r)$ of age-selected
samples of {\small SKID} haloes, based on the time at which the largest
progenitor reaches 50$\%$ of its peak mass.  Solid lines show $\xi(r)$
for all {\small SKID} haloes within each specified mass range.  The horizontal
and sloped dot-dashed lines correspond to $\xi(r)=1$, and
$\xi(r)=\left({r\over 3}\right)^{-1.75}$, respectively.}
\label{cfskidmages}
\end{center}
\end{figure}

One important difference between our $v_{c,peak}$-selected
haloes and the Gao \etal age dependence of clustering strength in
friends-of-friends haloes is that our sample has a large range in
masses whereas the Gao \etal study considered clustering at fixed
mass.  Because there is both a mass-age dependence and a
mass-clustering dependence in CDM models, it is useful to consider
what is the relation between age and clustering for {\small SKID} haloes at
fixed mass.  It should be noted that there are nontrivial dependencies
of {\small SKID} masses on environment; for example, less mass will be found to
be gravitationally self-bound in high density environments due mainly
to tidal stripping, but also affected to some degree by the addition of the
external potential in the computation of self-bound mass.  Thus, it is
not obvious that there should be a similar age dependence among {\small SKID}
haloes at fixed mass as there is for friends-of-friends haloes.  We
show in Fig. \ref{cfskidmages} that the age dependence of clustering
strength is indeed present for {\small SKID} haloes.  This effect has a strong
dependence on pair separation and on halo mass wherein the oldest
20$\%$ of the ${1.3-1.6 \times 10^{10} h^{-1} \msun}$ haloes have a
clustering amplitude approximately 10 times larger than that of the
the youngest 20$\%$ at scales of 0.5 $h^{-1}$Mpc.  At ${10^{11} h^{-1}
\msun}$, the effect is much weaker, consistent with little or no age
dependence, though our uncertainties are large in this higher mass
range due to the smaller number of haloes.

\section{discussion}
                   
We have shown that the clustering age dependence found for discrete
friends-of-friends haloes by Gao \etal is also present in a
mock galaxy catalog that consists of dark haloes plus satellites
selected by $v_{c,peak}$ to correspond to probable galaxy hosts.  This
provides strong evidence that galaxy clustering properties depend on
the assembly epoch of the dark matter hosts for a population that
includes field galaxies as well as galaxies within groups. 

It has been suggested by \eg Gao \etal that the age dependence of the
clustering amplitude may be a problem for the models of galaxy
formation (\eg Kauffmann, Nusser, \& Steinmetz 1997; Benson \etal
2000; Wechsler \etal 2001) and the models of galaxy clustering in the
halo model (\eg Seljak 2000; Cooray \& Sheth 2002; Berlind \etal 2003;
van den Bosch, Yang, \& Mo 2003) that assume that statistical galaxy
properties depend only upon halo mass.  
The result that the age
dependence on clustering is weaker among group members suggests that
any impact on these models will likely be strongest for the low mass
haloes that typically host $\simlt$1 luminous galaxy.

Though the clustering age dependence is very strong for the
extremes of the mock galaxy population, the age dependence is
generally weaker, and has a different scale dependence
than the colour dependence observed in recent large
surveys.  We expect that these qualitative differences are not sensitive
to the precise selection criteria of the simulated sample.   
The different behavior of the simulated
clustering age dependence suggests that
luminosity-weighted galaxy ages, \ie colour, do not trace halo age.
An apparent observed lack of correlation of stellar ages and halo ages
is evident by the general trend that massive (luminous) galaxies tend
to have old (red) stellar populations, in apparent contradiction to
the inverse mass-age relation present in hierarchical structure
formation.  This naively suggests that local environmental effects may
have a strong influence on galaxy stellar populations.  However, the
different merger rates of haloes in regions of different density are
also important because progenitor haloes of a given mass will have
been assembled earlier in more massive present day haloes, leading to
earlier star formation (\eg Mouri \& Taniguchi 2006; Neistein, van den
Bosch \& Dekel 2006).

In any case, ages and colours of galaxies are expected to be influenced
by a large number of astrophysical phenomena, including the
suppression of star formation by winds, AGN, or ram pressure stripping
(see \eg Berlind \etal 2005; Bower \etal 2005).  It may thus be
difficult to decrypt a halo age-clustering signal in the real
universe, as many of the influences on galaxy properties may depend on
mass, environment, or other parameters (see \eg Abbas \& Sheth 2006;
Cooray 2006).  However, there may still exist some correlation between
stellar population and halo age, if for example, major mergers trigger
major starbursts, or if the age of the oldest stars in a galaxy is
correlated with halo age. 
A recent 
semi-analytical study by Croton, Gao, \& White (2007) suggests that
that the clustering of group or cluster central galaxies 
should correlate with group host dark matter halo assembly age.
Some evidence for an
age-clustering trend of group haloes has recently been found by
observing that groups of similar mass, whose central galaxies have
more passive star formation, which may indicate earlier group
assembly, are more strongly clustered (Yang, Mo \& van den Bosch
2006).  See however, an apparently opposite relation found
by Berlind \etal (2006), who find that massive groups tend to be
less strongly clustered if they have redder central galaxies.  
To measure this effect in individual galaxies, one would
need to measure accurately star formation histories of galaxies hosted
by a narrow ranges of halo masses (see \eg Heavens \etal 2004 for
discussion of age measurements in SDSS stellar populations).
Comparisons between the clustering age dependence among simulated
haloes and the age dependence among galaxy stellar populations could
then provide clues to the physics of galaxy formation.

\subsection{Summary}
 
Within a high resolution cosmological dark matter simulation, we have
examined the clustering properties of a mock galaxy catalog selected
by $v_{c,peak}$ to match approximately the luminosity range  
and number density of observable galaxies.

$\bullet$ A strong clustering age dependence is found for mock galaxy
catalogs that include both central haloes and satellite haloes, and is
reflected in both spatial and in kinematic clustering measures.
It is caused primarily by 1) the age clustering relation for
discrete virialized haloes, acting on field mock galaxies, 
and 2) the contribution of group and cluster members, which
tend to be older, and are highly clustered due to their presence 
within massive dark matter hosts,
thereby increasing the tendency for old members of the full sample to be
highly clustered.

$\bullet$ The strength of the clustering age dependence implies that
it is likely to be manifested in real galaxies.  The clustering age
dependence is weaker than the clustering colour dependence in 2dF and
SDSS for pair separations less than $\sim$5 $h^{-1}$Mpc, 
and has a different scale dependence.  This means
that the observed clustering colour-dependence cannot be fully
explained by assuming that stellar population ages trace halo ages.
That is, one cannot simply assume that red galaxies lie in old haloes
and blue galaxies lie in young haloes.  The clustering colour
dependence must be influenced by additional processes that affect the
baryons.

$\bullet$ The clustering age dependence is weaker among group and
cluster mock catalog members than for the general galaxy population.

\section*{Acknowledgments}
DR has been supported by PPARC, and acknowledges early funding by a NASA
GSRP fellowship while at the University of Washington.  FG was
supported as a David E. Brooks Research Fellow, and was partially
supported by NSF grant AST-0098557 at the University of Washington. TQ
was partially supported by the NSF grant PHY-0205413.  We thank
Geraint Harker, Carlos Frenk, and Richard Bower for useful discussion.
Simulations were performed on the Origin 2000 at NCSA and NASA Ames,
the IBM SP4 at the Arctic Region Supercomputing Center (ARSC), and the
NASA Goddard HP/Compaq SC 45. We thank Chance Reschke for dedicated
support of our computing resources, much of which were graciously
donated by Intel.

{}

\label{lastpage}

\end{document}